# The Nuclear Spin Nanomagnet

V.L. Korenev

A. F. Ioffe Physical Technical Institute, St. Petersburg, 194021 Russia

*Linearly polarized light tuned slightly below the optical transition of the negatively charged exciton (trion) in a single quantum dot causes the spontaneous nuclear spin polarization (self-polarization) at a level close to 100 %. The effective magnetic field of spin-polarized nuclei shifts the optical transition energy close to resonance with photon energy. The resonantly enhanced Overhauser effect sustains the stability of the nuclear self-polarization even in the absence of spin polarization of the quantum dot electron. As a result the optically selected single quantum dot represents a tiny magnet with the ferromagnetic ordering of nuclear spins – the nuclear spin nanomagnet.*

The hyperfine interaction of nuclei with optically oriented electrons brings about a whole number of striking non-linear phenomena in bulk semiconductors. Among them are the optically governed magnetic hysteresis and the undamped auto-oscillations of polarization [1] persisting up to the liquid nitrogen temperature [2]. Magnetic hysteresis in the electron-nuclear spin system has been observed in nanostructures, too [3]. The most prominent consequence of the nonlinearity is the nuclear dynamic self-polarization (DSP) – spontaneous nuclear polarization close to 100 % in a weak (~1 mT) magnetic field in the absence of electron spin polarization. Although DSP was not found experimentally, its observation and examination may have profound consequences in the field of quantum information.

According to Dyakonov and Perel (DP) original idea [4] the DSP appears when the polarization of electron spins is maintained artificially (optically or electrically) equal to zero. The spontaneous nuclear polarization creates an effective hyperfine nuclear magnetic field that splits electron spin sublevels. The electron spin flow from up to down energy level due to the electron-nuclear flip-flop transitions (Overhauser effect) provides positive feedback, thus giving non-zero nuclear polarization. The DP mechanism requires very low temperature $T \sim A \sim 1$ K because the hyperfine constant $A \approx 1$ K is very small.

Korenev [5] proposed a different DSP scenario which is effective in nanostructures and takes place at elevated temperatures. It is based on the "coalescence" of spin levels of non-polarized quasiparticles: level crossing accelerates Overhauser effect. Spin sublevels of bright and dark excitons [5] approach each other in a spontaneous nuclear field, thus initiating the flip-flop transitions between the nearest exciton spin sublevels. The resonantly enhanced Overhauser effect provides positive feedback that maintains the nuclear self-polarization even in the absence of exciton spin polarization. Merkulov [6] considered the nonlinear exciton-nuclear quantum dot (QD) spin system pumped by circularly polarized light. He found the bistability of the exciton-nuclear spin system that has been observed recently in QD [7]. Recently the idea of Ref. [5] was used to predict the DSP in electronic transport through the double-dot system [8].

Here we show that the monochromatic linearly polarized light tuned slightly below the optical transition of the negatively charged exciton (trion) causes the spontaneous nuclear spin polarization (self-polarization) in a single quantum dot. The spontaneous effective magnetic field of spin-polarized nuclei shifts the trion optical transition energy toward resonance with photon energy. In turn, the resonantly enhanced Overhauser effect sustains the stability of nuclear polarization even in the absence of spin polarization of resident electron. As a result the optically selected single quantum dot represents a tiny magnet with the ferromagnetic ordering of nuclear spins – the nuclear spin nanomagnet.

Consider a singly charged with one electron QD of disc shape, with its normal pointing along z-axis. Electron states consist of two spin projections (spin up $|+1/2\rangle \equiv |\uparrow\rangle$ and spin down $|-1/2\rangle \equiv |\downarrow\rangle$). The QD is assumed to have a large number of nuclei ($N>>1$) each of which has spin $I$ and $(2I+1)$ states $|\mu\rangle$ with angular momentum projection $\mu = -I,...,+I$. The nuclei are not polarized in equilibrium. Assume that the nuclear polarization fluctuation $P_N$ larger than the equilibrium value $\sim I/\sqrt{N}$ appears along z direction. Polarized nuclei create an effective hyperfine magnetic field which splits the electron spin states by the $AIP_N$ value [1]. Energy conservation law suppresses the electron-nuclear flip-flop transitions. Below we show that the monochromatic linearly polarized light with energy $\hbar\omega$ slightly below the optical transition energy $E_T$ into the trion state initiates Overhauser effect and stabilizes the non-zero nuclear polarization. The optically excited trion consists of three particles – one hole and two electrons. The total angular momentum of trion $|\uparrow\downarrow\Uparrow\rangle$ $(|\uparrow\downarrow\Downarrow\rangle)$ is determined only by the heavy hole spin projections $|+3/2,h\rangle \equiv |\Uparrow\rangle$ and $|-3/2,h\rangle \equiv |\Downarrow\rangle$ onto z-axis. Unlike the single electron state, the trion state is doubly degenerated because the contact hyperfine interaction with hole is negligible, and electron spins are antiparallel.

The near-resonant ($\hbar\omega < E_T$) optically induced Overhauser effect is a second-order process (Fig.1). The bold arrow in Fig.1A shows the nuclear-spin-assisted absorption: the electron-nuclear flip-flop transition $|\uparrow,\mu\rangle \xrightarrow{hf} |\downarrow,\mu+1\rangle$ due to the hyperfine interaction (dashed line labeled hf) is followed by the photon-assisted (dashed line, ph) transition from the electron $|\downarrow\rangle$ state into the trion $|\uparrow\downarrow\Downarrow\rangle$ state [9]. At the end of transition the nuclear angular momentum is increased by 1. Participation of the photon satisfies the energy conservation law. Trion recombination through the much more efficient first-order spin conservative process (open arrow) emits a spontaneous photon $|\downarrow\uparrow\Downarrow,\mu+1\rangle \xrightarrow{ph} |\downarrow,\mu+1\rangle$, thus ending the optical cycle. As a result, the initial fluctuation $P_N$ is enhanced [10]. Another optical cycle (Fig.1A') increases the initial polarization, too: the first-order absorption of photon $|\uparrow,\mu\rangle \xrightarrow{ph} |\uparrow\downarrow\Uparrow,\mu\rangle$ is completed by the second-order nuclear-assisted emission $|\uparrow\downarrow\Uparrow,\mu\rangle \xrightarrow{ph} |\uparrow,\mu\rangle \xrightarrow{hf} |\downarrow,\mu+1\rangle$. Surely, there are competitive processes decreasing the nuclear spin polarization (Figs.1B, B'). Fig.1B shows the second-order absorption $|\downarrow,\mu+1\rangle \xrightarrow{hf} |\uparrow,\mu\rangle \xrightarrow{ph} |\uparrow\downarrow\Uparrow,\mu\rangle$ ended by the first-order emission into the $|\uparrow,\mu\rangle$ final state. Finally, the absorption $|\downarrow,\mu+1\rangle \xrightarrow{ph} |\downarrow\uparrow\Downarrow,\mu+1\rangle$ in Fig.1B' is completed by the second-order $|\downarrow\uparrow\Downarrow,\mu+1\rangle \xrightarrow{ph} |\downarrow,\mu+1\rangle \xrightarrow{hf} |\uparrow,\mu\rangle$ emission of photon. The key point here is that the probabilities of the optical cycles in Fig.1B, B' are less than those in Fig.1A,A' due to the energy conservation problem (the bold arrows in Fig.1B,B' do not reach the trion states). Therefore, the net result is the increase of the initial polarization $P_N$. In turn, the increased nuclear field shifts the transitions in Fig.1B, B' further out of (and the transitions in Fig.1A,A' close to) resonance. This is the essence of instability. The steady-state nuclear polarization is determined by the non-linear effects and can have values close to 100 %. This scheme differs from the previous DSP mechanism [5, 6, 8] based on the "coalescence" in energy of spin sublevels. Here it is the optical transition energy that is shifted toward resonance with the energy of photon. One should not also confuse this approach with the one based on the light-induced change of the level splitting [11]. Here the direct effect of laser power on the transition energy is negligible [10]. It is the spontaneous nuclear field that changes the level splitting. The Overhauser effect provides positive feedback, thus fixing the non-zero value of the nuclear field. Note that only the laser energy $\hbar\omega < E_T$ will produce DSP. The laser tuned above the trion resonance provides the negative feedback, thus preventing DSP. This asymmetry results from the positiveness of the hyperfine constant A>0. Both signs of $P_N$ are equally possible. Application of a small magnetic field (B~1mT) is useful for two reasons [4]: (i) to fix the sign of spontaneous polarization and (ii) to eliminate the effect of the nuclear dipole-dipole interaction destroying $P_N$.

Of course, the presented scheme is very approximate. For example, it does not take into account the uniform broadening of optical transitions. Rather strong broadening will smear the resonance. It means that the competitive transitions will prevent self-polarization and the quantitative study is necessary to get the conditions for DSP. The hyperfine interaction conserves the total spin of electron-nuclear system, so that the steady-state equation for Overhauser effect reads [1]

$$W^A_{\mu,\mu+1} n_\uparrow p_\mu + W^{A'}_{\mu,\mu+1} n_{+3/2} p_\mu = W^B_{\mu+1,\mu} n_\downarrow p_{\mu+1} + W^{B'}_{\mu+1,\mu} n_{-3/2} p_{\mu+1} \qquad (1)$$

where $n_\uparrow$ ($n_\downarrow$) and $n_{+3/2}$ ($n_{-3/2}$) are the average populations of spin up (down) electron and $|+3/2\rangle = |\uparrow\downarrow\Uparrow\rangle$ ($|-3/2\rangle = |\uparrow\downarrow\Downarrow\rangle$) trion states, $p_\mu$ – the number of QD nuclei with spin projection $\mu$. $W^{A(A')}_{\mu,\mu+1}$ and $W^{B(B')}_{\mu+1,\mu}$ are the probabilities of the A(A') and B(B') processes rising and decreasing the projection of nuclear spin by one (see Fig.1A(A') and Fig.1B(B')).

The nuclear self-polarization takes place when the resident electron spins are maintained non-polarized [1]. In our case the hole spin flip should be suppressed: the hole spin flip depletes the $|\uparrow\rangle$ state of the resident electron through the optical cycle $|\uparrow\rangle \xrightarrow{ph} |\uparrow\downarrow\Uparrow\rangle \xrightarrow{\text{hole spin flip}} |\uparrow\downarrow\Downarrow\rangle \xrightarrow{ph} |\downarrow\rangle$ [12]. This depletion eliminates both the Overhauser effect and the DSP (Figs.1A, A'). For the same reason we allow for the spin relaxation of resident electron. If temperature $T$ is larger than the nuclear field induced spin splitting $A \approx 1 K$ then the electron spin sublevels are equally populated [13]. In these conditions the resident electron is non-polarized, i.e. $n_\uparrow = n_\downarrow = 0.5$.

Although the emission-related Overhauser terms (the second terms on RHS and LHS of Eq.(1)) contain the small populations of the trion states (in spontaneous regime $n_{\pm 3/2} \ll n_{\uparrow(\downarrow)}$ [10]) they are comparable with the absorption-related terms (the first terms in both RHS and LHS of Eq.(1)) [14]. Indeed, compare Fig.1A and Fig.1A' and neglect the spin relaxation of holes that mixes the populations of $\pm 3/2$ trion states. Then in steady state conditions the absorption-induced and the emission-induced flip-flop rates are equal: $W^A_{\mu,\mu+1} n_\uparrow = W^{A'}_{\mu,\mu+1} n_{+3/2}$. Similarly one has $W^B_{\mu+1,\mu} n_\downarrow = W^{B'}_{\mu+1,\mu} n_{-3/2}$. As a result the Eq.(1) simplifies to $W^A_{\mu,\mu+1} n_\uparrow p_\mu = W^B_{\mu+1,\mu} n_\downarrow p_{\mu+1}$. Then

$$\frac{p_\mu}{p_{\mu+1}} = \frac{W^B_{\mu+1,\mu}}{W^A_{\mu,\mu+1}} \tag{2}$$

The probabilities $W^A_{\mu,\mu+1}$ and $W^B_{\mu+1,\mu}$ of optically induced flip-flop transitions are proportional to the uniformly broadened densities of the final trion states (see Appendix)

$$W^A_{\mu,\mu+1} \propto \frac{\Gamma}{\Gamma^2 + (E_T - \hbar\omega - E_\uparrow)^2}, \quad W^B_{\mu+1,\mu} \propto \frac{\Gamma}{\Gamma^2 + (E_T - \hbar\omega - E_\downarrow)^2} \tag{3}$$

Here parameter $\Gamma$ is the halfwidth of the uniformly broadened absorption line having Lorenzian shape [15]; $E_\uparrow = +AIP_N/2$ and $E_\downarrow = -AIP_N/2$ are the energies of spin up and spin down initial (resident electron) state. With the use of Eqs.(2,3) the nuclear polarization degree $P_N$ can be expressed in terms of the Brillouin function

$$P_N = B_I(x), \quad \text{where } x = I \cdot \ln\left(\frac{(\Delta + AIP_N/2)^2 + \Gamma^2}{(\Delta - AIP_N/2)^2 + \Gamma^2}\right) \tag{4}$$

Here $\Delta = E_T - \hbar\omega$ gives the detuning from the trion optical transition energy $E_T$. Eq.(4) gives the solution of the problem. It has trivial solution $P_N=0$, corresponding to unordered state. Under certain conditions the other two roots of opposite signs appear. They correspond to the spontaneous nuclear polarization along or opposite to z-axis. Near the critical points the polarization value is small ($P_N \ll 1$), and the analytical solution of Eq.(4) is possible. Expanding the Brillouin function up to the cubic term we have

$$P_N = \pm \frac{2}{A \cdot I} \sqrt{\frac{E_0 \cdot \Delta - \Delta^2 - \Gamma^2}{b(\Delta, \Gamma)}} \quad (5)$$

where $E_0 = 2AI(I+1)/3$ and parameter $b(\Delta, \Gamma) = 1 + \frac{4}{15}(2 \cdot I + 3)(2 \cdot I - 1)\frac{\Delta^2}{\Delta^2 + \Gamma^2}$ are positive.

One can see from Eq.(5) that the nontrivial behavior takes place for positive detuning $\Delta > 0$, i.e. *below* the trion resonance in accord with previous qualitative discussion. The ordered state exists in the limited region of detuning $\Delta$ (where radicand expression is positive), i.e. between

$$\Delta_- = \frac{E_0}{2} - \sqrt{\left(\frac{E_0}{2}\right)^2 - \Gamma^2} \quad \text{and} \quad \Delta_+ = \frac{E_0}{2} + \sqrt{\left(\frac{E_0}{2}\right)^2 - \Gamma^2} \quad (6)$$

The physics of the first transition point $\Delta_-$ was discussed qualitatively above. The second transition point $\Delta_+$ at large detuning appears because both levels are out of resonance with photon energy, thus giving approximately the same (small) probabilities of light absorption and suppressing DSP. Shaded area in Fig.2 shows the region of non-trivial solutions of Eq.(4) in $\Delta$-$\Gamma$ axes. We conclude that the nontrivial solutions exist only if the level broadening is not large: $E_0 > 2\Gamma$. Let us check this condition for the GaAs-type natural quantum dot in which *A=90 μeV, I=3/2, Γ=30 μeV* [16]. We have $E_0$=225 μeV that is 7.5 times larger than *Γ*, so that the condition $E_0 > 2\Gamma$ can be easily satisfied in experimental conditions.

Strictly speaking, Eq.(4) tells nothing about the stability of steady states. The fact that the Overhauser effect represents dynamical problem was realized long ago [17]. To analyze the stability of the solutions, the Eq.(4) should be replaced with the dynamical equation. We shall use the equation similar to that in Ref. [17]

$$\frac{dP_N}{dt} = -\frac{1}{T_{1e}}[P_N - B_I(x(P_N))] \quad (7)$$

where $T_{1e}$ characterizes the dynamic nuclear polarization time. Eq.(7) togehter with Eq. (4) extends the previous approach to the dynamic case with the polarization $P_N$ being dynamical variable [18]. In a steady state Eq.(7) reduces to the Eq.(4). The linearization of the Eq.(7) near the steady state point $P_N$=0 shows that trivial solution is unstable at $E_0 > 2 \cdot \Gamma$ within the detuning range $[\Delta_-, \Delta_+]$. The state with spontaneous nuclear polarization is stable within the shaded area in Fig.2 while the unordered state is stable in the whitened area.

Solid line in Fig.3a shows the stable solution of numerically calculated Eqs (4,7) near the first transition point $\Delta_- \approx 4\,\mu eV$ for *A=90 μeV, I=3/2, Γ=30 μeV*. One can see (dashed line) that approximate formula (5) works well near the transition point as is expected. Fig.3b demonstrates the behavior near the second transition point $\Delta_+ \approx 220\mu eV$. Approximate formula (5) gives practically the same result as numerical calculation (solid and dashed lines coincide). The overall dependence $P_N(\Delta)$ is calculated numerically and shown in Fig.3c. The maximum nuclear polarization value (about 97 % due to the small contribution of B,B' processes) is achieved when the detuning energy $\Delta_{max} \approx 70\mu eV$. This corresponds to the strictly resonant excitation of the low energy component of the doublet split by almost totally polarized nuclei, i.e. $\Delta_{max} = AIP_N^{max}/2 \approx AI/2 \approx 70\,\mu eV$. A rather high nuclear polarization value can be reached in the wide range of detuning: $P_N$>75 % for Δ up to 150 *μeV*.

Throughout the paper we neglected the spin relaxation of holes and allowed it for electrons. Quantitative analysis (Appendix) shows that the condition $\gamma\tau_h \gg 1 + G_r\tau_s$ favors DSP ($G_r$ - generation rate of trions, $\gamma$ is the spontaneous emission rate of trion, $\tau_h, \tau_s$ are the spin relaxation times of hole and electron, respectively). The suppression of the hole spin flip was proved experimentally in Ref. [19] which found the parameter $\gamma\tau_h \geq 10$ up to T=30K. Possible source of the electron spin relaxation is the particle (and spin) exchange between the QD electron and n-doped substrate [20]. The substrate contains a reservoir of non-polarized electrons. Although the average QD charge is equal to one it is renewed via the spin exchange with the reservoir. The renewal time has a meaning of time τ$_s$ and should satisfy the above condition. One should keep in mind, however, that this time should not be too short, because each electron jump stimulates the leakage of nuclear spin at a rate $1/T_L \approx 1/\tau_s \cdot q$, where q – the probability of the electron-nuclear flip-flop transition accompanying the electron jump. The leakage rate should be less than the optically induced Overhauser effect rate $1/T_{1e} \approx G_r \cdot q$ that gives the condition $G_r\tau_s>1$ [21]. Therefore the most favorable condition for DSP is $\gamma\tau_h >> G_r\tau_s > 1$. In the spontaneous regime ($\gamma >> G_r$) it can be satisfied even at $\tau_h \leq \tau_s$.

Therefore, the experiment on selective nuclear dynamic self-polarization is to be carried out within the temperature range 5-30 K. A weak (~1 mT) magnetic field along z-axis is necessary to suppress the effect of nuclear dipole-dipole interaction. Linearly polarized laser excites the individual QD with laser energy being tuned below the ground state trion line by $\geq 2\Gamma$ [22]. The laser intensity should be strong enough to initiate Overhauser effect but should not induce the level broadening. The DSP will manifest itself in the appearance (with characteristic time $T_{1e}$) of the splitting of the trion luminescence by $\approx 70$ $\mu eV$ in spontaneous Overhauser field. Another possible experimental manifestation of DSP is the rotation of light polarization plane by nuclear magnetic field [23].

Author is grateful to K.V. Kavokin and I.A. Merkulov for fruitful discussions. This work was supported by RFBR 05-02-17796, Program of RAS.

# APPENDIX

## 1. Rate equations and steady state populations of states in a single QD.

The system of equations describing the populations of resident electron $n_{\uparrow(\downarrow)}$ and trion $n_{\pm 3/2}$ states in the QD in the presence of spin relaxation of electrons with time $\tau_s$ and holes with time $\tau_h$ and the energy level scheme are

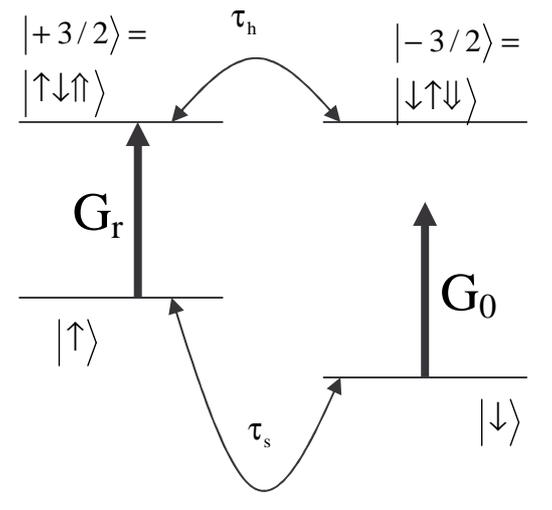

$$\begin{cases} \dfrac{dn_{+3/2}}{dt} = G_r n_\uparrow - \gamma n_{+3/2} - \dfrac{n_{+3/2} - n_{-3/2}}{2\tau_h} \\[4pt] \dfrac{dn_{-3/2}}{dt} = G_0 n_\downarrow - \gamma n_{-3/2} - \dfrac{n_{-3/2} - n_{+3/2}}{2\tau_h} \\[4pt] \dfrac{dn_\uparrow}{dt} = -G_r n_\uparrow + \gamma n_{+3/2} - \dfrac{n_\uparrow - n_\downarrow}{2\tau_s} \\[4pt] \dfrac{dn_\downarrow}{dt} = -G_0 n_\downarrow + \gamma n_{-3/2} - \dfrac{n_\downarrow - n_\uparrow}{2\tau_s} \\[4pt] n_\uparrow + n_\downarrow + n_{+3/2} + n_{-3/2} = 1 \end{cases}$$

Here $G_r$ ($G_0$) is the generation rate of optical transition close to (far from) resonance; $\gamma$ – radiative decay rate. The last equation means that the QD is singly charged. In this scheme the electron and hole spin generation/relaxation rates are assumed to be much faster than the electron-nuclear flip-flop transitions, so that at this stage we neglect small corrections coming from the hyperfine interaction. We restrict ourselves by the low power density regime when stimulated optical transitions are negligible in compare with spontaneous emission: $\gamma \gg G_r, G_0$. In this case the steady state solutions are

$$n_{+3/2} = \frac{G_0 + G_r + 2G_r \gamma \tau_h + 2G_0 G_r \tau_s}{2\gamma[2 + 2\gamma\tau_h + (G_0 + G_r)\tau_s]}; \quad n_{-3/2} = \frac{G_0 + G_r + 2G_0 \gamma \tau_h + 2G_0 G_r \tau_s}{2\gamma[2 + 2\gamma\tau_h + (G_0 + G_r)\tau_s]} \tag{A1}$$

$$n_\uparrow = \frac{1 + \gamma\tau_h + G_0 \tau_s}{[2 + 2\gamma\tau_h + (G_0 + G_r)\tau_s]}; \quad n_\downarrow = \frac{1 + \gamma\tau_h + G_r \tau_s}{[2 + 2\gamma\tau_h + (G_0 + G_r)\tau_s]} \tag{A2}$$

One can see that $n_\uparrow + n_\downarrow = 1$ and $n_{\pm 3/2} \ll 1$ as it should be. Three parameters ($\gamma \tau_h$, $G_0 \tau_s$, $G_r \tau_s$) are important. Generally $G_r \geq G_0$. If $\gamma \tau_h \gg G_r \tau_s$ then $n_\uparrow = n_\downarrow = 1/2$, i.e. the resident electron is non-polarized.

2. **Calculation of probabilities of A, A', B, B' processes.**

The optically induced Overhauser effect is a two-step process that should be calculated in the second-order perturbation theory. The A-process (absorption-induced Overhauser effect, Fig.1A) increases the nuclear projection momentum by one. The probability per second of the optically induced $|\uparrow,\mu\rangle \to |\downarrow,\mu+1\rangle$ flip-flop transition of the resident electron with anyone of the N quantum dot nuclei is

$$W^A_{\mu,\mu+1} = \frac{2\pi}{\hbar} \frac{\left|\langle\downarrow\uparrow\Downarrow|\hat{V}_{ph}|\downarrow\rangle\right|^2 \left|\langle\downarrow,\mu+1|\hat{V}_{hf}|\uparrow,\mu\rangle\right|^2}{(E_\uparrow - E_\downarrow)^2} \frac{\Gamma/\pi}{\Gamma^2 + (E_T - \hbar\omega - E_\uparrow)^2} \quad (A3)$$

where $\langle\downarrow\uparrow\Downarrow|\hat{V}_{ph}|\downarrow\rangle$ - the matrix element of optical transition between $|\downarrow\rangle$ electron and $|\downarrow\uparrow\Downarrow\rangle$ trion states; the matrix element of contact hyperfine interaction is $\langle\downarrow,\mu+1|\hat{V}_{hf}|\uparrow,\mu\rangle$. The last multiplier in Eq.(A3) represents the uniformly broadened density of final states with the halfwidth $\Gamma$. Eq.(A3) can be rewritten as

$$W^A_{\mu,\mu+1} = G_r \cdot q_\mu \quad (A4)$$

where the optical generation rate into the $|\downarrow\uparrow\Downarrow\rangle$ trion state

$$G_r = \frac{2\pi}{\hbar} \left|\langle\downarrow\uparrow\Downarrow|\hat{V}_{ph}|\downarrow\rangle\right|^2 \frac{\Gamma/\pi}{\Gamma^2 + (E_T - \hbar\omega - E_\uparrow)^2}, \quad (A5)$$

parameter $q_\mu = \dfrac{\left|\langle\downarrow,\mu+1|\hat{V}_{hf}|\uparrow,\mu\rangle\right|^2}{(E_\uparrow - E_\downarrow)^2}$ is the probability of flip-flop transition accompanying the absorption process. It can be evaluated similar to the Ref. [24] taking into account the contact Hamiltonian $\hat{V}_{hf} = \sum_{i=1}^{N} A_i \vec{I}_i \cdot \vec{S}$. Here $A_i$ characterizes the hyperfine coupling strength between the electron and the i-th nucleus ($A_i \sim A/N$, because only $N$-th "part" of electron is located at a

given nucleus). For our purposes it is enough to estimate it. Hyperfine interaction of the electron with a given nucleus mixes the electron spin sublevels with amplitude $\sim \frac{A/N}{(E_\uparrow - E_\downarrow)}$, where $E_\uparrow - E_\downarrow = AIP_N \sim A$ - the splitting of the electron spin states due to the effective hyperfine field of polarized nuclei. Thus the probability of flip-flop transition $q_\mu \sim N\left(\frac{A/N}{A}\right)^2 \sim \frac{1}{N}$ is very small because $N>>1$.

Another absorption-induced Overhauser effect (B-process, Fig.1B) decreases the nuclear momentum projection by one

$$W^B_{\mu+1,\mu} = G_0 \cdot q_\mu \tag{A6}$$

where

$$G_0 = \frac{2\pi}{\hbar} \left|\langle \uparrow\downarrow\Uparrow | \hat{V}_{ph} | \uparrow \rangle\right|^2 \frac{\Gamma/\pi}{\Gamma^2 + (E_T - \hbar\omega - E_\downarrow)^2} \tag{A7}$$

gives the generation rate into the $|\uparrow\downarrow\Uparrow\rangle$ trion state. For the linearly polarized light excitation $\left|\langle \uparrow\downarrow\Uparrow | \hat{V}_{ph} | \uparrow \rangle\right|^2 = \left|\langle \downarrow\uparrow\Downarrow | \hat{V}_{ph} | \downarrow \rangle\right|^2$. Thus the Eq.(A7) looks similar to the Eq.(A5) excepting for the energy denominator due to the energy difference between the initial spin up and down states.

Emission related Overhauser effect gives somewhat different results. The A'-process (Fig.1A') increases the nuclear spin by one with probability

$$W^{A'}_{\mu,\mu+1} = \frac{2\pi}{\hbar} \sum_{\substack{photon \\ modes}} \frac{\left|\langle \downarrow, \mu+1 | \hat{V}_{hf} | \uparrow, \mu \rangle\right|^2 \left|\langle \uparrow | \hat{V}_{ph} | \uparrow\downarrow\Uparrow \rangle\right|^2}{(E_\uparrow - E_\downarrow)^2} \frac{\Gamma/\pi}{\Gamma^2 + (E_T - \hbar\omega - E_\downarrow)^2} \tag{A8}$$

where the summation goes over all modes of emitted photons. Denote emission rate as $\gamma = \frac{2\pi}{\hbar} \sum \left|\langle \uparrow | \hat{V}_{ph} | \uparrow\downarrow\Uparrow \rangle\right|^2 \frac{\Gamma/\pi}{\Gamma^2 + (E_T - \hbar\omega - E_\downarrow)^2} = \frac{2\pi}{\hbar} |V|^2_{ph} \rho_{ph}$, where the density $\rho_{ph}$ of photon states varies slowly with frequency. Thus we obtain

$$W^{A'}_{\mu,\mu+1} = \gamma \cdot q_\mu \tag{A9}$$

Emission-induced Overhauser effect for the B'-process (Fig.1B') can be found similarly

$$W^{B'}_{\mu+1,\mu} = \gamma \cdot q_\mu \tag{A10}$$

Eqs. (A4, A6, A9, A10) show that additional assisting process (absorption or emission of photon) initiates the Overhauser effect. Participation of photon helps to satisfy the energy conservation law. The rates of the optically induced flip/flop transitions are much smaller than those for the assisting processes with no spin flips due to the small q-factor.

With the use of Eqs.(A1,A2), (A4,A9) and (A6,A10) one can see that in the absence of hole spin relaxation $\gamma\tau_h \to \infty$ the absorption-induced flip-flop rate is equal to the emission-induced flip-flop rate: $W^A_{\mu,\mu+1}n_\uparrow = W^{A'}_{\mu,\mu+1}n_{+3/2} = G_r q_\mu/2$ and $W^B_{\mu+1,\mu}n_\downarrow = W^{B'}_{\mu+1,\mu}n_{-3/2} = G_0 q_\mu/2$. This justifies the Eq.(2) of the main text.

## 3. Nuclear dynamic polarization. The most favorable conditions for DSP

To calculate the steady state nuclear polarization we use (i) the Eq.(1) for the Overhauser effect; (ii) Eqs. (A1, A2) for the electron and trion state populations; (iii) Eqs. (A4, A6, A9, A10) for the flip-flop transition probabilities. We have

$$\frac{p_\mu}{p_{\mu+1}} = \frac{W^B_{\mu+1,\mu}n_\downarrow + W^{B'}_{\mu+1,\mu}n_{-3/2}}{W^A_{\mu,\mu+1}n_\uparrow + W^{A'}_{\mu,\mu+1}n_{+3/2}} = \frac{G_0 n_\downarrow + \gamma n_{-3/2}}{G_r n_\uparrow + \gamma n_{+3/2}} = \frac{(G_0+G_r)/2 + 2G_0 G_r \tau_s + G_0(1+2\gamma\tau_h)}{(G_0+G_r)/2 + 2G_0 G_r \tau_s + G_r(1+2\gamma\tau_h)} \tag{A11}$$

The initial fluctuation of nuclear polarization $P_N$ increases when $p_\mu < p_{\mu+1}$. Such a tendency is clearly seen from Eq.(A11) because $G_r > G_0$. The most favorable condition for the DSP requires the ratio $p_\mu/p_{\mu+1}$ to be as small as possible. It comes from the consideration of limiting cases. In resonance the absorption of the off-resonant component is absent ($G_0 \approx 0$). The Eq. (A11) reduces to $1/(3+4\gamma\tau_h)$, thus giving $\gamma\tau_h \gg 1$. On the other hand, for the $G_r \geq G_0$ case the most favorable condition is $\gamma\tau_h \gg G_r\tau_s$. Combining the limiting cases, we get

$$\gamma\tau_h \gg 1 + G_r\tau_s \tag{A12}$$

It means the blocking of hole spin flip but finite electron spin relaxation time (actually the electron and hole spin relaxation times can be comparable $\tau_s \sim \tau_h$ because $\gamma \gg G_r$). In these conditions the resident electrons are non-polarized $n_\uparrow = n_\downarrow = 0.5$ (see Eq.(A2)) in agreement with qualitative discussion given in the main text. In the limit (A12)

$$\frac{p_\mu}{p_{\mu+1}} = \frac{G_0}{G_r} + \frac{1}{4\gamma\tau_h} = \frac{(\Delta - AIP_N/2)^2 + \Gamma^2}{(\Delta + AIP_N/2)^2 + \Gamma^2} + \frac{1}{4\gamma\tau_h} \qquad (A13)$$

To derive the last equality we used Eqs. (A5, A7), and $E_{\uparrow,\downarrow} = \pm AIP_N/2$. In resonance $(\Delta = AIP_N/2)$ the ratio (A13) is the smallest when

$$\gamma\tau_h \gg \left(\frac{AI}{2\Gamma}\right)^2 \qquad (A14)$$

This condition may be more strict than (A12). However, in practice (A=90 $\mu$eV, I=3/2, $\Gamma = 30\,\mu eV$) we have $(AI/2\Gamma)^2 \approx 5$, so that the conditions (A12) and (A14) are comparable. Neglecting the second term on RHS of Eq.(A13) and expressing Eq.(A13) via the Brillouin function we reproduce the Eq. (4) of the main text.

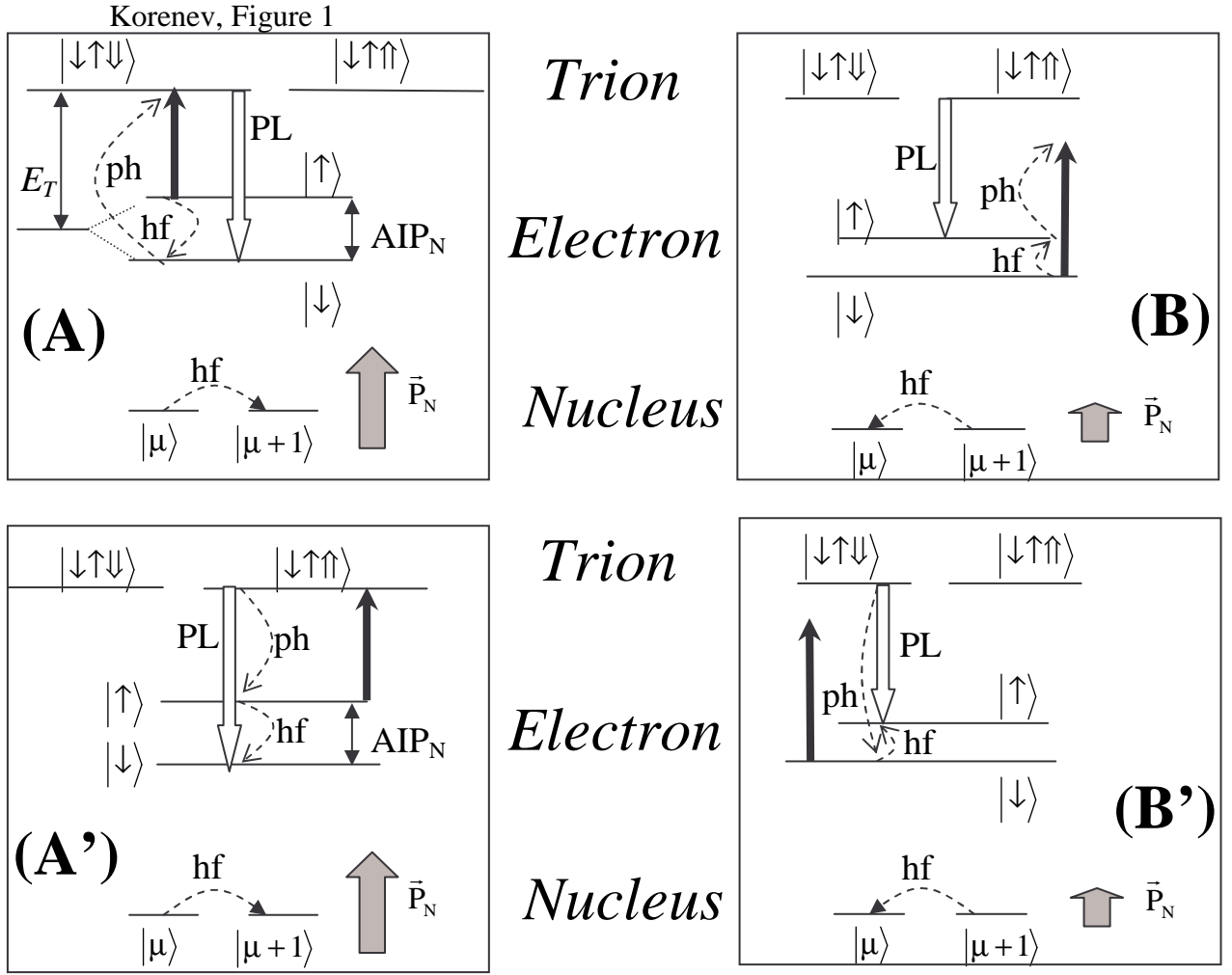

Optically induced Overhauser effect. Optical cycles (A-A') increase the nuclear spin projection by one: (A) Electron-nuclear flip-flop transition $|\uparrow,\mu\rangle \xrightarrow{hf} |\downarrow,\mu+1\rangle$ is followed by the photon absorption $|\downarrow,\mu+1\rangle \xrightarrow{ph} |\downarrow\uparrow\Downarrow,\mu+1\rangle$; the first-order photon emission $|\downarrow\uparrow\Downarrow,\mu+1\rangle \xrightarrow{ph} |\downarrow,\mu+1\rangle$ ends the cycle. (A') Photon absorption $|\uparrow,\mu\rangle \xrightarrow{ph} |\uparrow\downarrow\Uparrow,\mu\rangle$ starts the cycle. The second-order process $|\uparrow\downarrow\Uparrow,\mu\rangle \xrightarrow{ph} |\uparrow,\mu\rangle \xrightarrow{hf} |\downarrow,\mu+1\rangle$ ends the cycle. Optical cycles (B-B') decrease the nuclear polarization: (B) the second-order absorption $|\downarrow,\mu+1\rangle \xrightarrow{hf} |\uparrow,\mu\rangle \xrightarrow{ph} |\uparrow\downarrow\Uparrow,\mu\rangle$ starts the cycle ended by the $|\uparrow\downarrow\Uparrow,\mu\rangle \xrightarrow{ph} |\uparrow,\mu\rangle$ first-order emission. (B') The absorption $|\downarrow,\mu+1\rangle \xrightarrow{ph} |\downarrow\uparrow\Downarrow,\mu+1\rangle$ is completed by the second-order $|\downarrow\uparrow\Downarrow,\mu+1\rangle \xrightarrow{ph} |\downarrow,\mu+1\rangle \xrightarrow{hf} |\uparrow,\mu\rangle$ emission of photon. Dashed arrows represent single virtual processes in second-order processes.

Korenev, Figure 2

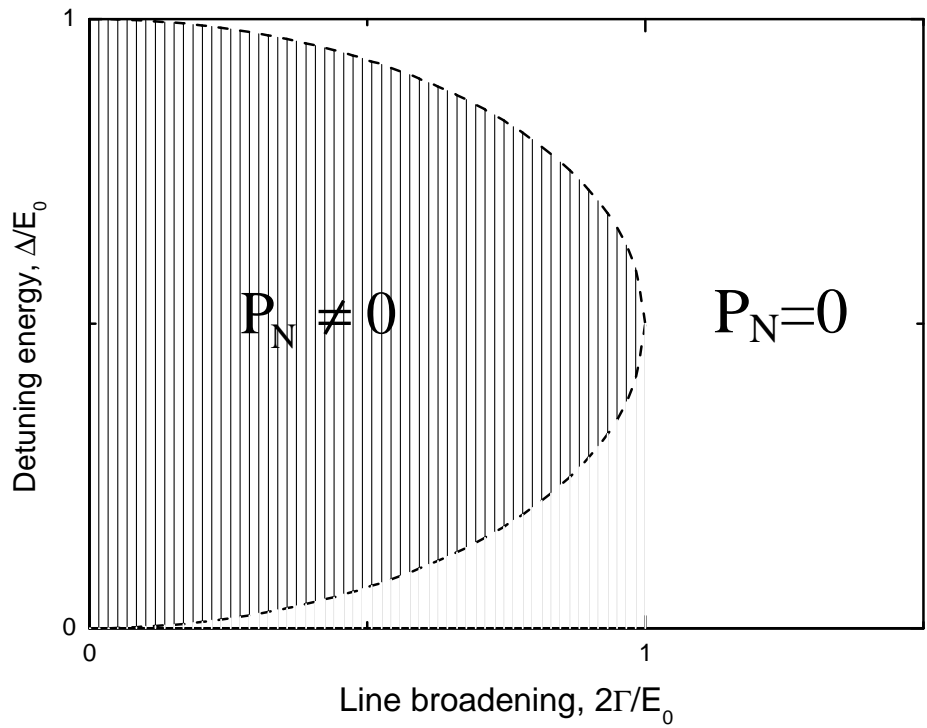

"Phase" diagram of the nuclear spin system in QD. Shaded area represents the stable state with ferromagnetic order of nuclear spins – dynamic self-polarization regime. Whitened area shows unordered stable state with $P_N=0$.

Korenev , Figure 3

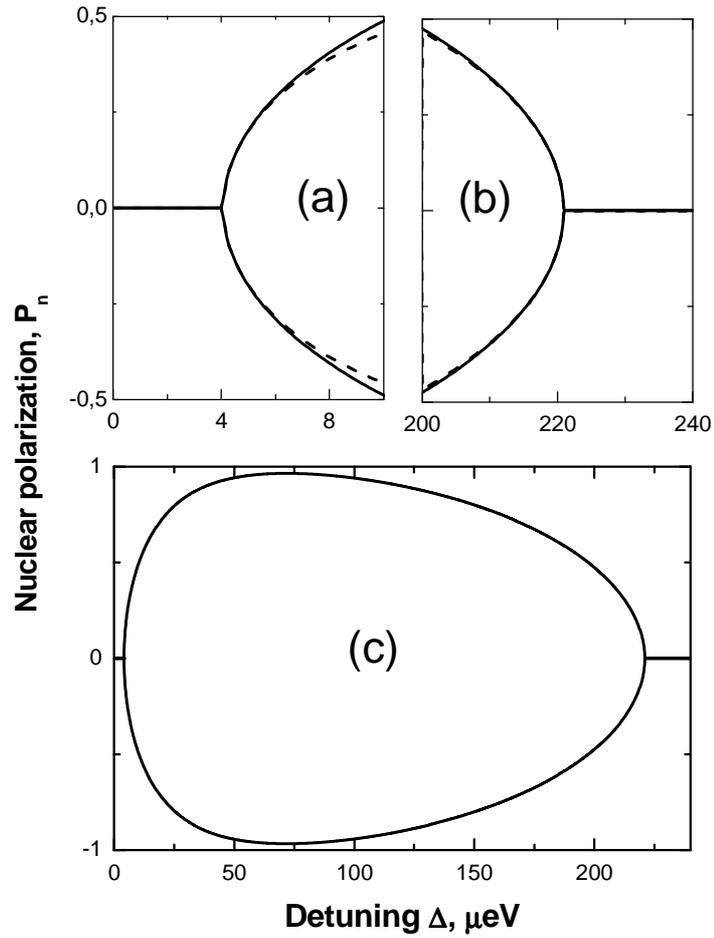

Stable solutions $P_N(\Delta)$ calculated numerically according to Eqs (4,7) (solid line) and analytically by Eq.(5) (dashed lines) with the use of parameters *A=90 μeV, I=3/2, Γ=30 μeV* near (a) the first transition point $\Delta_- \approx 4\,\mu eV$; (b) the second transition point $\Delta_+ \approx 220\,\mu eV$. (c) the overall dependence $P_N(\Delta)$ is calculated numerically by Eqs.(4,7).